\documentclass[a4paper]{book}
\usepackage{epsf,nano2cmr}
\usepackage{graphicx,epsfig,color,bm}
\begin{document}

\pnum{}
\ttitle{Optical bistability in artificial composite nanoscale molecules:\\
Towards all optical processing at the nanoscale}

\tauthor{{\em A.~V.~Malyshev}, V.~A.~Malyshev}

\ptitle{Optical bistability in artificial composite nanoscale molecules:\\
Towards all optical processing at the nanoscale}

\pauthor{{\em A.~V.~Malyshev}$^{1,2}$,
V.~A.~Malyshev$^{3}$}

\affil{$^{1}$~Universidad Complutense de Madrid, Spain\\
$^{2}$~\FTI\\
$^{3}$~Zernike Institute for Advanced Materials, University of Groningen,
The Netherlands
}

\begin{abstract} {Optical response of artificial composite nanoscale
molecules comprising a closely spaced noble metal nanoparticle and a
semiconductor quantum dot have been studied theoretically. We consider a
system composed of an Au particle and CdSe or CdSe/ZnSe quantum dot and
predict optical bistability and hysteresis in its response, which suggests
various applications, in particular, all-optical processing and optical
memory.} \end{abstract}

\begindc

\index{Malyshev A. V.}
\index{Malyshev V. A.}

\section*{Introduction} Arrays of metallic nano-particles (often referred to
as plasmonic arrays), are widely recognized as potential building blocks for
nanoscale optical circuits~\cite{Quinten98,Brongersma00,Malyshev08}.
Recently, a number of papers reported fascinating properties of small
clusters of closely spaced semiconductor quantum dots (SQD) and metallic
nano-particles (MNP)~\cite{Zhang06,Govorov06,Artuso08,Sadeghi09}. Fano
resonances~\cite{Zhang06,Artuso08}, bistability in the absorption
spectrum~\cite{Artuso08}, and meta- ``molecular''
resonances~\cite{Sadeghi09} have been predicted. When such systems are
excited optically, the dipole moment of the excitonic transition in the SQD
generates additional electric field at the MNP, which is superposed with the
external field, while the induced dipole moment of the MNP generates
an additional electric field at the SQD, providing a feedback. Thus, the
presence of the MNP leads to a self-action of the optical transition
dipole moment, which can give rise to a variety of new optical properties.
In particular, if the coupling between two particles is strong the
self-action can be large enough to result in optical bistability.

\begin{figure}[t]
%\leavevmode
\centering{
%\epsfbox[width=0.45\textwidth]{malyshev01.eps}
\includegraphics[width=0.3\columnwidth]{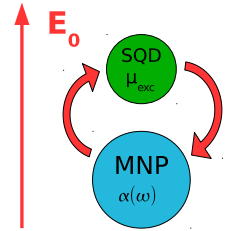}
}
\caption[]{ Schematics of a semiconductor quantum dot - metallic
nanoparticle hybrid system embedded into a homogeneous dielectric background
with permeability $\varepsilon_b$ and subjected to an external field with
the amplitude ${{\bf E}_0}$. }
\label{Schematics}
\end{figure}

We consider a CdSe or CdSe/ZnSe spherical semiconductor quantum dot
and a Au nano-sphere, the simplest artificial diatomic nano-molecule
(see the schematic view of the system in Fig.~\ref{Schematics}). We
demonstrate that optical bistability and hysteresis can be observed
in the system. The two stable states of the systems have different
polarizations, providing a possibility to store information in the
form of the system polarization. We argue also that in addition to
the traditional way of switching by the amplitude of the driving
electric field, the state of such artificial diatomic molecule can
be switched by the polarization of the field with respect to the
molecule axis. The fact that both the SQD and the MNP can sustain
high electric fields suggests such possible applications of the
artificial molecule as an all-optical switch and optical memory at
nano-scale.

\section*{Formalism}

We assume that the SQD-MNP system is embedded in a dielectric host
with the permeability $\varepsilon_b$ and driven by the external
electric field with the amplitude ${\bf E}_0$ and frequency
$\omega$. The SQD is treated quantum mechanically (as a two level
system) within the framework of the Maxwell-Bloch equations while
the MNP is treated classically; the response of the MNP is described
by its frequency dependent polarizability within the point dipole
approximation. The rotating wave approximation is used throughout
the paper, so that all time dependent quantities represent
amplitudes of the corresponding signals, the set of equations for
which reads:
%
%\begin{eqnarray}
%\dot Z = - \gamma\,(Z+1)-\frac{1}{2}
%\left[\,
%\Omega\,R^{*}+\Omega^{*}R
%\,\right]
%\\
%\nonumber
%\dot R = -(i\,\Delta+\Gamma)\, R+\Omega\,Z
%\qquad\qquad\;\;\,\ ,
%\label{MB}
%\end{eqnarray}
%
\begin{equation}
\dot Z = - \gamma\,(Z+1)-\frac{1}{2} \left(\,
\Omega\,R^{*}+\Omega^{*}R \,\right) \ ,
\label{MB1}
\end{equation}
\begin{equation}
\dot R = -(i\,\Delta+\Gamma)\, R+\Omega\,Z \ , \label{MB2}
\end{equation}
where $Z=\rho_{11}-\rho_{00}$ is the population difference, $R$ is
the amplitude of the off-diagonal density matrix element
$\rho_{10}$, $\gamma$ and $\Gamma$ are relaxation rates, $\Delta$ is
the detuning of the SQD resonance and $\Omega={\bm\mu\bf E}/\hbar$
is the Rabi frequency ($\bm\mu$ is the optical transition dipole
moment of the SQD). The total electric field $\bf E$ at the SQD is
the superposition of the external field ${\bf E}_0$ and the
scattered field produced by the MNP:
\begin{equation}
\label{E_SQD} {\bf E}=\frac{1}{\varepsilon_{\mathrm{s}}^{\prime}}
\left( {\bf E}_0+\frac{ {\bar{\bf S}}\,{\bf P}_{\mathrm{MNP}}
}{\varepsilon_b\,d^3} \right) \ .
\end{equation}
Here, $\varepsilon_{\mathrm{s}}^{\prime}=
3\varepsilon_{\mathrm{b}}/(\varepsilon_{\mathrm{s}} +
2\varepsilon_{\mathrm{b}})$, $\varepsilon_{\mathrm{s}}$ is the SQD
dielectric constant, ${\bar{\bf S}} = \mathrm{diag}(-1,-1,2)$ is the
angular part of the dipole field Green tensor, $d$ is the SQD-MNP
center-to-center distance and ${\bf P}_\mathrm{MNP}$ is the induced
dipole moment of the MNP:
\begin{equation}
\label{P_MNP}
{\bf P}_{\mathrm{MNP}}=\alpha(\omega)\,a^3
\left(
{\bf E}_0+\frac{{\bar{\bf S}}\,{\bf P}_{\mathrm{SQD}}}{\varepsilon_b\,d^3}
\right) \ ,
\end{equation}
where $\alpha(\omega)a^3$ is the classical frequency dependent
polarizability of the MNP, $a$ being its radius and
$\alpha(\omega)=[\varepsilon_M(\omega)-\varepsilon_b]/[\varepsilon_M(\omega)+2\varepsilon_b]$
with $\varepsilon_M(\omega)$ being the dielectric function of the metal
while ${\bf P}_{\mathrm{SQD}}=(-i/2)R\,{\bm \mu}$ is the SQD dipole moment
amplitude. The total electric field in the SQD is therefore given by
\begin{equation}
\label{E_SQD1}
{\bf E}= \frac{1}{\varepsilon_{\mathrm{s}}^{\prime}}
\left[
{\bf 1}+\frac{\alpha(\omega)\,a^3}{\varepsilon_b\,d^3}\,{\bar{\bf S}}
\right]{{\bf E}_0}
+
\frac{\alpha(\omega)\,a^3}{\varepsilon_{\mathrm{s}}^{\prime}
\varepsilon_b\,d^6}\,{\bar{\bf S}}^2\,{\bf P}_{\mathrm{SQD}}\ .
\end{equation}
The Rabi frequency $\Omega=\Omega_0-i\,G\,R$ entering
Eq.~(\ref{MB1}) and~(\ref{MB2}) contains the part corresponding to
the renormalized external field, $\Omega_0$ [the first term in
Eq.~(\ref{E_SQD1})], and the self action [the second term in
Eq.~(\ref{E_SQD1})]. The steady-state equation for the population
difference $Z$ reads:
\begin{equation}
\label{Z}
\frac{|\Omega_0|^2}{\gamma\,\Gamma}=
-\frac{Z+1}{Z}\,
\frac{|\Gamma+i\,(\Delta+G\,Z)|^2}
{\Gamma^2}
\end{equation}
For some set of parameters and values of the driving fields this
equation can have three real solutions, only two of which turn up to
be stable.

\section*{Results}

Hereafter we consider spherical CdSe/ZnSe SQD and Au MNP and use the
following set of parameters: $\hbar\omega=2.36$ eV (which
corresponds to optical transition in 3.3 nm SQD),
$\varepsilon_s=6.2$, $a=10$ nm, $d=17$ nm, $\varepsilon_b=1$,
$1/\gamma=0.8$ ns, $1/\Gamma=0.3$ ns, $\Delta=0$. We use tabulated
dielectric function of gold~\cite{Johnson72} to calculate the
polarizability of the MNP.

\begin{figure}[ht]
\begin{center}
\includegraphics[width=0.45\textwidth]{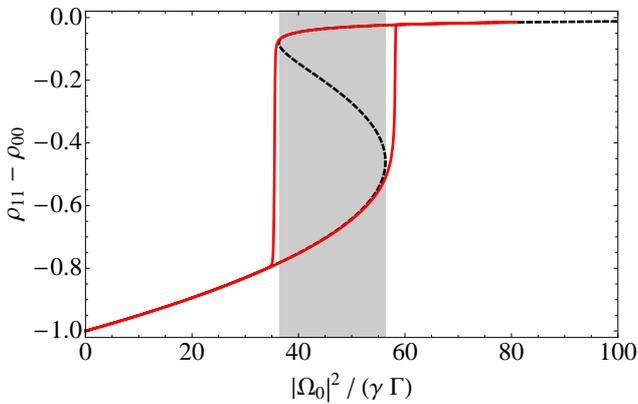}
\end{center}
\caption{
Dependence of the population difference of the SQD on the renormalized
external field $\Omega_0$ manifesting optical bistability (shaded region).
Dashed line -- the three-valued solution of Eq.~(\ref{Z}). Solid line --
solution of Eq.~(\ref{Z}) with time dependent external field sweeping back
and forth across the bistability region.
}
\label{bist}
\end{figure}

Dashed line in Fig.~\ref{bist} shows the solution to Eq.~(\ref{Z}),
the dependence of the steady state population difference on the
renormalized external field $\Omega_0$ (the field is parallel to the
system axis). Shaded region shows the range of the external field
for which the system can have three different states. Only the upper
and lower branches of this solution are stable, resulting in optical
bistability. To observe the two branches, we sweep the external
field back and forth across the bistability region. Solid line in
Fig.~\ref{bist} shows the result of the corresponding calculation.
The figure demonstrates that when the field is increasing (from 0),
the population difference follows the lower stable branch until the
critical field amplitude is reached. Further, the system switches
abruptly to its other stable state (upper branch). When we sweep the
field back the upper-to-lower branch switch occurs at a different
critical field, forming the typical hysteresis loop.
%
%The slower the sweeping speed, the closer the loop gets to the steady state
%solution.

\begin{figure}[t]
\begin{center}
\includegraphics[width=0.4\textwidth]{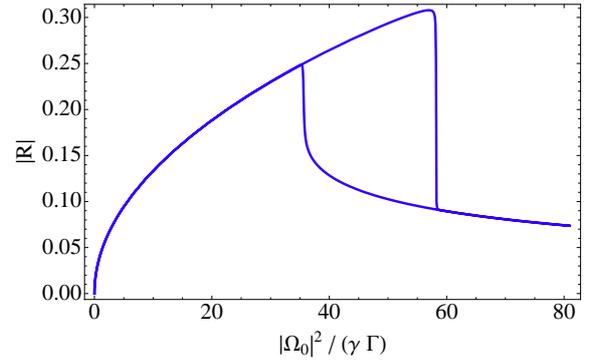}
\end{center}
\caption{
Dependence of the dipole moment amplitude $|R|$ of the SQD on the
renormalized external field $\Omega_0$, manifesting optical
bistability in the polarization of the SQD.
}
\label{hyst}
\end{figure}

The bistability of the population of the SQD results in the bistability of
SQD and MNP polarizations. In Fig.~\ref{hyst}, we present the hysteresis
loop of the polarization of the SQD calculated for the same parameters as in
Fig.~\ref{bist}. The two stable states of the SQD population correspond to
two different polarization, providing the possibility for storage of
information by the system polarization.

Finally, we point out that the specific feature of the SQD-MNP
system is the existence of the symmetry axis; this allows us to
drive the system not only by the amplitude of the external field,
but also by the orientation of its polarization with respect to the
axis. These are new properties which can be useful for applications.

\section*{Summary}
We investigated theoretically the optical response of a hybrid
``artificial'' molecule composed of a semiconductor quantum dot, modeled as
a two-level system and a metal nanoparticle, considered classically, which
are coupled by the dipole-dipole interaction. The interaction results in a
self-action of the SQD via MNP, leading to the dependence of the SQD optical
transition frequency on the population, which provide a feedback mechanism
resulting in several fascinating effects. Thus, in the strong coupling
regime, we found that the system can manifest bistability and optical
hysteresis, as well as switching of the polarization of both SQD and MNP by
the incoming field. Such switching can be achieved not only by the
traditional amplitude control but also by the polarization of the incoming
field with respect to the system axis; the latter being very promising for
optical memory applications at the nanoscale.

\ack A. V. M. acknowledges the program Ram\'on y Cajal (Ministerio
de Ciencia e Innovaci{\'o}n de Espa{\~n}a) for support and the
University of Groningen for hospitality.


\begin{thebibliography}{8}
\itemsep-2pt

\bibitem{Quinten98} M. Quinten {\em et al.}, Opt. Lett. {\bf 23}, 1331 (1998).

\bibitem{Brongersma00} M. L. Brongersma, J. W. Hartman, and H. A. Atwater,
    Phys. Rev. B {\bf 62}, R16356 (2000).

\bibitem{Malyshev08} A. V. Malyshev and V. A. Malyshev, {\em Nano
    Lett.} {\bf 8}, 2369 (2008).

\bibitem{Zhang06} W. Zhang, A. O. Govorov, and G. W. Bryant,  {\em Phys. Rev.
    Lett.} {\bf 97}, 146804 (2006).

\bibitem{Govorov06} A. O. Govorov {\em et al.}, {\em Nano Lett.} {\bf 6}, 984 (2006).

\bibitem{Artuso08} R. D. Artuso and G. W. Bryant, {\em Nano Lett.} {\bf 8},
    2106 (2008).

\bibitem{Sadeghi09} S. M. Sadeghi, {\em Phys. Rev. B} {\bf 79}, 233309 (2009).

\bibitem{Johnson72} P. B. Johnson and R. W. Christy, {\em Phys. Rev. B} {\bf
6}, 4370 (1972)

%\bibitem{Benedict90}M. G. Benedict, A. I. Zaitsev, V. A. Malyshev,
%         and E. D. Trifonov, {\em Opt. Spektrosk.} {\bf 68}, 812 (1990)
%     [Opt. Spectrosc. {\bf 68} 473 (1990)].

%\bibitem{Benedict88} M. G. Benedict, V. A. Malyshev, E. D.
%    Trifonov, and A. I. Zaitsev, {\em Phys. Rev. A} {\bf 43}, 3845 (1991).

%\bibitem{Malyshev95} V. Malyshev and P. Moreno, Phys. Rev. B {\bf 51},
%    14587 (1995).

%\bibitem{Malyshev98} V. A. Malyshev, H. Glaeske, and K.-H. Feller,
%    {\em Phys. Rev. B} {\bf 58}, 1496 (1998).

%\bibitem{Friedberg89} R. Friedberg, S. R. Hartmann, and J. T. Manassah,
%   {\em Phys. Rev. A} {\bf 39}, 3444, 1989.


\end{thebibliography}
\end{document}